\documentclass[10pt,twocolumn,amssym,amsmath,bm]{./IEEEtran_v15}
\usepackage{epsfig}

\makeatletter
\def\ifundefined{\@ifundefined}
\makeatother

\begin{document}

\title{\Large \textbf{
Parallelizing the Keldysh formalism for strongly correlated electrons}
}

    \author{\large J.~K.~Freericks\\
    \normalsize \textit{ Department of Physics}\\
    \normalsize \textit{ Georgetown University}\\
    \normalsize  \textit{Washington, DC 20057}\\
    \normalsize  \textit{Email: freericks@physics.georgetown.edu}
    \and
    \large V.~M.~Turkowski\\
    \normalsize \textit{Department of Physics}\\
    \normalsize \textit{Georgetown University}\\
    \normalsize \textit{Washington, DC 20057}\\
    \normalsize \textit{Email: turk@physics.georgetown.edu}
    \and
    \large V.~Zlati\'c\\
    \normalsize \textit{Institute of Physics}\\
    \normalsize \textit{Bijenicka c. 46, P. O. B. 304}\\
    \normalsize \textit{10000 Zagreb, Croatia}\\
    \normalsize \textit{Email: zlatic@ifs.hr}}

\ifundefined{IEEEtransversionmajor}{%
   
   %
   \newlength{\IEEEilabelindent}
   \newlength{\IEEEilabelindentA}
   \newlength{\IEEEilabelindentB}
   \newlength{\IEEEelabelindent}
   \newlength{\IEEEdlabelindent}
   \newlength{\labelindent}
   \newlength{\IEEEiednormlabelsep}
   \newlength{\IEEEiedmathlabelsep}
   \newlength{\IEEEiedtopsep}

   \providecommand{\IEEElabelindentfactori}{1.0}
   \providecommand{\IEEElabelindentfactorii}{0.75}
   \providecommand{\IEEElabelindentfactoriii}{0.0}
   \providecommand{\IEEElabelindentfactoriv}{0.0}
   \providecommand{\IEEElabelindentfactorv}{0.0}
   \providecommand{\IEEElabelindentfactorvi}{0.0}
   \providecommand{\labelindentfactor}{1.0}
   
   \providecommand{\iedlistdecl}{\relax}
   \providecommand{\calcleftmargin}[1]{
                   \setlength{\leftmargin}{#1}
                   \addtolength{\leftmargin}{\labelwidth}
                   \addtolength{\leftmargin}{\labelsep}}
   \providecommand{\setlabelwidth}[1]{
                   \settowidth{\labelwidth}{#1}} 
   \providecommand{\usemathlabelsep}{\relax}
   \providecommand{\iedlabeljustifyl}{\relax}
   \providecommand{\iedlabeljustifyc}{\relax}
   \providecommand{\iedlabeljustifyr}{\relax}
 
   \newif\ifnocalcleftmargin
   \nocalcleftmarginfalse

   \newif\ifnolabelindentfactor
   \nolabelindentfactorfalse
   
   \newif\ifcenterfigcaptions
   \centerfigcaptionsfalse
   
   \let\OLDitemize\itemize
   \let\OLDenumerate\enumerate
   \let\OLDdescription\description
   
   \renewcommand{\itemize}[1][\relax]{\OLDitemize}
   \renewcommand{\enumerate}[1][\relax]{\OLDenumerate}
   \renewcommand{\description}[1][\relax]{\OLDdescription}

   \providecommand{\pubid}[1]{\relax}
   \providecommand{\pubidadjcol}{\relax}
   \providecommand{\specialpapernotice}[1]{\relax}
   \providecommand{\overrideIEEEmargins}{\relax}
   
   \let\CMPARstart\PARstart 
   
   \let\OLDappendix\appendix
   \renewcommand{\appendix}[1][\relax]{\OLDappendix}
   
   \newif\ifuseRomanappendices
   \useRomanappendicestrue
   
   \let\OLDbiography\biography
   \let\OLDendbiography\endbiography
   \renewcommand{\biography}[2][\relax]{\OLDbiography{#2}}
   \renewcommand{\endbiography}{\OLDendbiography}
   
   \markboth{A Test for IEEEtran.cls--- {\tiny \bfseries
   [Running Older Class]}}{Shell: A Test for IEEEtran.cls}}{
  
   }

%
%

\maketitle
     \thispagestyle{empty}
     \pagestyle{empty}


\begin{abstract}
Nonequilibrium quantum mechanics can be solved with the Keldysh formalism,
which evolves the quantum mechanical states forward in time in the 
presence of a time-dependent field, and then evolves them backward in time,
undoing the effect of the time-dependent field.  The Feynman path integral
over the Keldysh contour is employed to calculate the strongly
correlated Green's function.  We examine the accuracy of this procedure for
the simplest problem that requires a nonequilibrium formulation: the
$f$-electron spectral function of the spinless Falicov-Kimball model.
\end{abstract}


\section{Introduction}
\PARstart{E}{lectrons} are correlated when the Coulomb repulsion
between them is strong enough that it plays a significant role in determining
the motion of the electrons through the crystal.  Correlated electrons
are of interest to the military, because their properties
(metallic/insulating/magnetic/superconducting/etc.) can be easily 
tuned by changing
pressure, chemical composition, irradiation with electromagnetic fields,
and so on, and form the basis of many so-called smart materials and
devices. In addition, a large number of materials
of interest to the military (like Plutonium) have strong electron correlations

Our research problem involves understanding strongly correlated
materials when they are placed under intense electromagnetic fields that
can drive them out of equilibrium creating interesting dynamical and
relaxational effects (examples include intense pulsed laser irradiation
or interacting with 
large amplitude microwaves). Our main focus is to the Josephson junction 
device~\cite{josephson}
(a sandwich of a superconductor-barrier-superconductor which has the potential
for ultra high speed digital electronics~\cite{rsfq}),
but the principles can be applied to a large number of different
devices.  The many-body formalism for nonequilibrium
problems is solved by a Feynman path integral over the so-called Keldysh
contour~\cite{keldysh,rammer_smith}
which involves an evolution forward in time as the external fields
are turned on, evolution out to a long time, then an inverse evolution backward
in time where the fields are turned off.  Solving the Feynman path integral
requires evaluating finite-sized determinants of discretized matrices that
represent the continuous matrix operator along the contour.  We typically
require the determinant of approximately  500 general complex matrices with
sizes up to about $2100\times 2100$ for a production run.  This computational
effort is easily parallelized.  The numerical solutions suffer from a
discretization error that gets worse as the temperature is lowered,
and accurate calculations require a careful extrapolation
with different discretization sizes to the limit where the discretization
goes to zero.  Here we benchmark the numerics by solving for
the $f$-electron spectral function of the spinless Falicov-Kimball 
model~\cite{falicov_kimball}.

The Falicov-Kimball model is the simplest model of electron correlations
(and the problem we investigate is the simplest nontrivial Keldysh problem).
It possesses two types of electrons: conduction electrons, which can hop
to any of their nearest neighbors and localized ($f$) electrons which are
localized on the lattice sites.  There is a Coulomb repulsion $U$ between
conduction electrons and localized electrons on the same lattice site.
As $U$ increases, the conduction band splits into two bands with an
energy gap in between, and undergoes the so-called Mott metal-insulator
transition.
In this contribution, we restrict ourselves and all formulas to the case
of half filling, where half of the lattice sites are occupied by the
conduction electrons and half by the localized electrons. In this case,
the Hamiltonian becomes~\cite{falicov_kimball}
\begin{eqnarray}
\mathcal{H}&=&-\frac{t^*}{\sqrt{Z}}\sum_{\langle ij \rangle} (c^\dagger_ic_j
+c^\dagger_jc_i)
+U\sum_i c^\dagger_ic_if^\dagger_if_i\nonumber\\
&-&\frac{U}{2}\sum_i(c^\dagger_ic_i
+f^\dagger_if_i),
\label{eq: ham_def}
\end{eqnarray}
where $c^\dagger_i$ ($c_i$) creates (destroys) a conduction electron at
site $i$, $f^\dagger_i$ ($f_i$) creates (destroys) a localized electron at
site $i$, $U$ is the interaction strength, and $t^*$ is the hopping integral.
The symbol $Z$ represents the number of nearest neighbors, and 
$\langle i j\rangle$ denotes a sum over all nearest neighbor pairs. 
The first term is the kinetic
energy of the conduction electrons, the second term is the Coulomb repulsion,
and the third term is the chemical potential times the filling for both
electrons. 

We solve this problem on an infinite coordination 
number~\cite{metzner_vollhardt} $(Z\rightarrow \infty)$
Bethe lattice, which has a noninteracting density of states (DOS) that is
a semicircle
\begin{equation}
\rho(\epsilon)=\frac{1}{2\pi t^{*2}}\sqrt{4t^{*2}-\epsilon^2}.
\label{eq: bethe_dos}
\end{equation}
The noninteracting
bandwidth is $4t^*$. We choose $t^*$ as our energy unit and set it equal to 1. 
See Ref.~\cite{freericks_review} for a review of the
equilibrium and linear response solutions.

\section{Formalism}

We start with an examination of the retarded local Green's function for the
conduction electrons, defined to be
\begin{equation}
G_{jj}(t)=-i\theta(t) \textrm{Tr} \langle e^{-\beta\mathcal{H}}
\{ c_j(t),c^\dagger_j(0)\}_+\rangle/\mathcal{Z},
\label{eq: g_cond_def}
\end{equation}
with $\theta(t)$ the unit step function, \textrm{Tr} denoting the trace over
all conduction electron and localized electron states of the lattice, 
$\beta=1/T$, 
$c_j(t)=\exp(it\mathcal{H})c_j\exp(-it\mathcal{H})$, the braces denote
an anticommutator, and $\mathcal{Z}=\textrm{Tr}\langle \exp(-\beta\mathcal{H})
\rangle$. We chose to evaluate the local Green's function at site $j$, but 
it is the same at every site when there is no long range order.
In the limit of infinite coordination number, we find that the many-body
self energy for the Green's function becomes local~\cite{metzner_vollhardt}, 
and the many-body problem
for the lattice can be mapped onto the many-body problem for an
impurity in a time-dependent field with a self-consistency relation to the
lattice~\cite{brandt_mielsch}.  The 
Fourier transform of the retarded Green's function can be
determined (on the Bethe lattice) by solving a simple cubic 
equation~\cite{hubbard,vandongen}
\begin{equation}
G^3(\omega)-2\omega G^2(\omega)+(1+\omega^2-\frac{U^2}{4})G(\omega)
-\omega=0,
\label{eq: cubic}
\end{equation}
where one must choose the (causal) physical solution determined by the root 
with a negative imaginary part (when the imaginary part is nonzero) and by
continuity when real. The conduction electron DOS is defined by 
$A(\omega)=-\textrm{Im}G(\omega)/\pi$, the electronic self energy 
(on the Bethe lattice) satisfies
\begin{equation}
\Sigma(\omega)=\omega+\frac{U}{2}-G(\omega)-\frac{1}{G(\omega)},
\label{eq: sigma}
\end{equation}
and the dynamical mean field $\lambda(\omega)$ is defined to satisfy
\begin{equation}
G(\omega)=\frac{1}{\omega+\frac{U}{2}-\lambda(\omega)-\Sigma(\omega)}
\label{eq: lambda}
\end{equation}
which can be thought of as the Fourier transform of the
time-dependent field for the impurity problem
(indeed the dynamical mean-field theory approach is to construct an
impurity problem in a time-dependent $\lambda$ field and then adjust the
field until the Green's functions for the impurity are equal to the local 
Green's functions for the lattice~\cite{brandt_mielsch}).

\begin{figure}
\epsfxsize=3.1in
\centerline{\epsffile{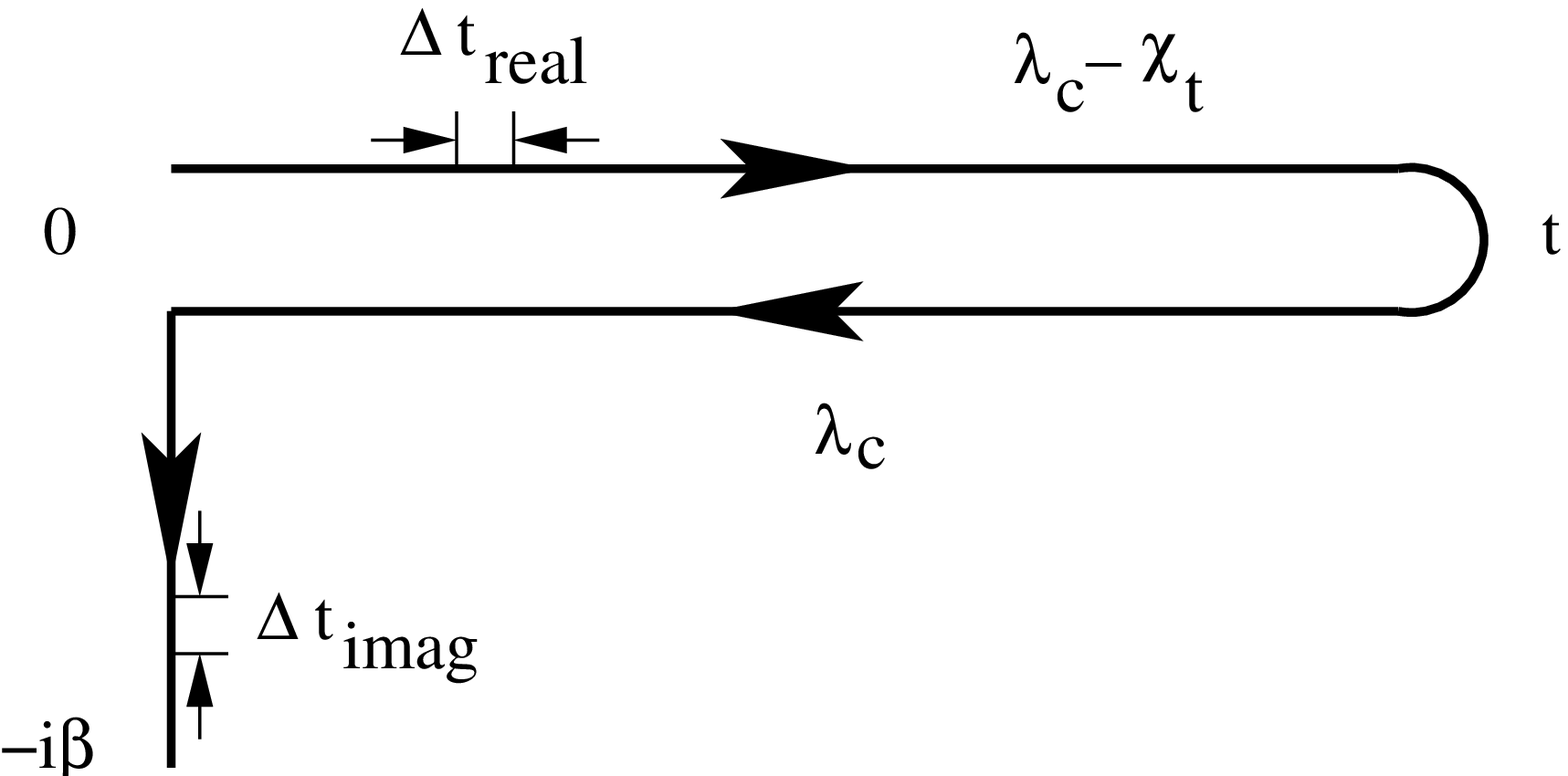}}
\caption{\label{fig: contour} Keldysh contour for evaluating the $f$-electron
Green's function at time $t$.  The contour runs from $\bar t=0$ to $\bar t =t$,
then back from $\bar t =t$ to $\bar t=0$ and finally goes
along the imaginary axis
down to  $\bar t= -i\beta$.  The field $\lambda_c-\chi_t$ is active on the
forward branch, and the field $\lambda_c$ is active over the backward branch
and the imaginary branch.  When we discretize the matrix operator over the 
Keldysh contour, we evaluate the integrals via a rectangular (midpoint) 
summation with a step size of $\Delta t_{\textrm{real}}$ on the real axis 
and $\Delta t_{\textrm{imag}}$ on the imaginary axis. We choose 
$\Delta t_{\textrm{imag}}=0.05$ and vary $\Delta t_{\textrm{real}}$ from 0.1 to
0.0125 in our calculations.  We typically use no more than 1000 time steps on
the outward branch of the contour.}
\end{figure}

In order to calculate the $f$-electron Green's function, we must first start
with the impurity problem, whose Hamiltonian is the same as the lattice
Hamiltonian, but there is no hopping term, since it is restricted to the
single site of the impurity.  The hopping of the conduction electrons is 
mimicked by the time dependent $\lambda$ field which destroys a conduction
electron at time 0 and creates a conduction electron at time $t$ with
``strength'' $\lambda(t)$.
The (greater) $f$-electron Green's function is then defined by
\begin{equation}
G_f^>(t)=-\textrm{Tr}\langle e^{-\beta \mathcal{H}_{imp}}S_c(\lambda)f(t)
f^\dagger(0)\rangle/\mathcal{Z}_{imp},
\label{eq: green_f_def}
\end{equation}
with $f(t)=\exp(it\mathcal{H}_{imp})f\exp(-it\mathcal{H}_{imp})$ and
the evolution operator given by
\begin{equation}
S_c(\lambda)=\mathcal{T}_c\exp\left [ \int_cd\bar t\int_c d\bar t^\prime
c^\dagger(\bar t)\lambda_c(\bar t, \bar t^\prime )c(\bar t^\prime )\right ] .
\label{eq: s_lambda_c}
\end{equation}
The time-ordering is along the Keldysh contour (see Fig.~\ref{fig: contour}), 
and the contour-ordered 
dynamical mean field is found from a Fourier transform of $\lambda(\omega)$
\begin{eqnarray}
\lambda_c(\bar t, \bar t^\prime )&=&-\frac{1}{\pi}\int_{-\infty}^\infty
d\omega \textrm{Im} \lambda(\omega)\exp[-i\omega(\bar t -\bar t^\prime )]\nonumber\\
&\times&[f_{FD}(\omega)-\theta_c(\bar t - \bar t^\prime )],
\label{eq: lambda_c}
\end{eqnarray}
where $f_{FD}(\omega)=1/[1+\exp(\beta\omega)]$ is the Fermi-Dirac distribution
and $\theta_c(\bar t - \bar t^\prime )=0$ if $\bar t^\prime$ is in front of
$\bar t$ on the contour $c$ and 1 if it is behind.

This Green's function can be solved directly by evaluating a Feynman
path integral over the Keldysh contour to 
yield~\cite{brandt_urbanek,freericks_review}
\begin{eqnarray}
G_f^>(t)&=&-\frac{e^{iUt/2}+e^{\beta U/2-iUt/2}}{\mathcal{Z}_{imp}}
\label{eq: g_f_t}\\
&\times&\textrm{Det}\left [ \delta_c(\bar t - \bar t^\prime)+
\int_c d\bar t^{\prime\prime}g^{aux}(\bar t, \bar t^{\prime\prime})\lambda
(\bar t^{\prime\prime},\bar t^\prime)\right ]\nonumber
\end{eqnarray}
which involves the determinant of a continuous matrix operator (note the path 
integral is the time-ordered product along the integration contour,
which is the Keldysh contour here).  The function
$g^{aux}$ appears in Table~\ref{table: g_aux}. It is the Green's function for
a noninteracting Fermion that evolves in a 
time-dependent field $\chi_t(\bar t,\bar t^\prime)
=-iU\theta_c(t-\bar t)\delta_c(\bar t - \bar t^\prime )$, which arises from
the time dependence of the localized 
electron~\cite{brandt_urbanek,freericks_review}.  In this sense,
one has the fields $\lambda_c -\chi_t$ acting on the forward branch of the
Keldysh contour, and the field $\lambda_c$ acting on the backward and imaginary
branches of the contour. The (equilibrium) greater Green's function
satisfies a spectral formula with the DOS and the Fermi-Dirac distribution
function (because the equilibrium distribution function is known)
\begin{equation}
G_f^>(t)=\int_{-\infty}^\infty d\omega e^{-i\omega t}[f_{FD}(\omega)-1]
A_f(\omega).
\label{eq: f_greater_spectral}
\end{equation}
Using the fact that the greater Green's function satisfies $G_f^>(t)=
G_f^{>*}(-t)$ and the fact that the DOS at half filling is an even function
of $\omega$ (due to particle-hole symmetry) yields the final equation
for the $f$-electron DOS~\cite{brandt_urbanek,freericks_review}
\begin{equation}
A_f(\omega)=-\frac{2}{\pi}\int_0^\infty dt \textrm{Re}\{G_f^>(t)\}
\cos(\omega t).
\label{eq: f_spect_final}
\end{equation}

\begin{table}[h]
\caption{$g^{aux}(\bar t,\bar t^\prime)$ for different orderings of
$t$, $\bar t$, and $\bar t^\prime$ along the contour $c$. The symbol
$\bar\xi_0$ satisfies $\bar\xi_0=1/[1+\exp(iUt-\beta U/2)]$.
\label{table: g_aux}}
\begin{tabular}{ll}
Green's function value&domain\\
\hline
$\bar\xi_0\exp[iU(\bar t-\bar t^\prime )/2]$& $t<\bar t<\bar t^\prime$\\
$\bar\xi_0\exp[iUt-iU(\bar t +\bar t^\prime)/2]$
& $\bar t< t<\bar t^\prime$\\
$\bar\xi_0\exp[-iU(\bar t-\bar t^\prime )/2]$& $\bar t<\bar t^\prime<t$\\
$(\bar\xi_0-1)\exp[iU(\bar t-\bar t^\prime)/2]$& $t<\bar t^\prime<\bar t$\\
$(\bar\xi_0-1)\exp[-iUt+iU(\bar t+\bar t^\prime)/2]$
& $\bar t^\prime<t<\bar t$\\
$(\bar\xi_0-1)\exp[-iU(\bar t-\bar t^\prime)/2]$
& $\bar t^\prime<\bar t<t$\\
\hline
\end{tabular}
\end{table}

In addition to computing the Green's function on the real frequency axis, one
can also compute it on the imaginary (Matsubara) frequency axis, at the
Matsubara frequencies $i\omega_n=i\pi T (2n+1)$.  In this case, one can
formulate the expressions as the determinant of a discrete matrix, so
there is no error associated with discretizing a continuous matrix 
operator~\cite{brandt_urbanek,freericks_review}.
Since the Matsubara Green's function can also be expressed as an integral over
the DOS
\begin{equation}
G_f(i\omega_n)=\int d\omega \frac{1}{i\omega_n-\omega}A_f(\omega),
\label{eq: g_mats}
\end{equation}
we have an independent way to verify the accuracy of the DOS by comparing the
integral formula for $G(i\omega_n)$ with the result directly calculated on
the imaginary axis.  Usually the accuracy is worst for the lowest Matsubara
frequency.

There also are a number of moments and properties of the DOS that can be tested.
First, the DOS is always nonnegative---negative values of the DOS for some 
region of frequency indicate an error in the calculations.  Second, one
can work out explicit values for the first three moments.  At half filling, 
these satisfy
\begin{eqnarray}
\int d\omega A_f(\omega)&=&1,\label{eq: moment1}\\
\int d\omega A_f(\omega)\omega f_{FD}(\omega)&=&U(\langle c^\dagger c 
f^\dagger f \rangle -\frac{1}{4}),\label{eq: moment2}\\
\int d\omega A_f(\omega)\omega^2&=&\frac{U^2}{4}.
\label{eq: moment3}
\end{eqnarray}
In practice we add a small shift (always less than 0.006 in magnitude) to
the DOS in order to satisfy the unit weight sum rule [Eq.~(\ref{eq: moment1})]. 
Then we check the
accuracy of the other two sum rules [by independently calculating the
correlation function on the right hand side of
Eq.~(\ref{eq: moment2})].  At $t=0$, we can use the moments in
Eq.~(\ref{eq: moment1}--\ref{eq: moment3}) to show that 
\begin{eqnarray}
G_f^>(0)&=&-\frac{1}{2} \label{eq: gf_0},\\
\frac{d}{dt}G_f^>(0)&=&-iU(\langle c^\dagger c f^\dagger f\rangle -\frac{1}{4})
\label{eq: d1_gf_0},\\
\frac{d^2}{dt^2}G_f^>(0)&=&\frac{U^2}{4}\label{eq: d2_gf_0}.
\end{eqnarray}
Thus the first derivative depends on temperature, but is purely imaginary,
while the second derivative is real
and independent of $T$.  Hence we expect $\textrm{Re}
G_f^>(t)$ to depend weakly on $T$ for small times, and to show stronger
dependence at large times.

\section{Computational Algorithm}

The main effort of this nonequilibrium calculation is to compute the
$f$-electron Green's function, which requires the determinant of
the continuous matrix operator in Eq.~(\ref{eq: g_f_t}).  
To calculate this, we must first 
decide on a discretization along the contour to evaluate line integrals
over the contour (see Fig.~\ref{fig: contour}).  Our choice is to use
a step size of $\Delta t_{\textrm{real}}$ on the real axis and
$\Delta t_{\textrm{imag}}=0.05$ along the imaginary axis. We use a midpoint
rectangular integration rule, evaluating the function at the midpoint of 
each discretized piece of the contour, for the approximation to the line
integral.  Hence
\begin{equation}
\int_c d\bar t h(\bar t)=\sum_j W_j h(\bar t_j),
\label{eq: integral sum}
\end{equation}
with the weight function $W_j$ satisfying $W_j=\pm\Delta t_{\textrm{real}}$ on
the real axis (positive on the forward branch and negative on the
backward branch) and $W_j=-i\Delta t_{\textrm{imag}}$ on the imaginary axis.
The times where the function is evaluated are $t_j=(j-1/2)
\Delta t_{\textrm{real}}$ on the forward branch, $t_j=(2j_{\textrm{max}}-j+1/2)
\Delta t_{\textrm{real}}$ on the backward branch, and 
$t_j=i(2j_{\textrm{max}}-j+1/2)\Delta t_{\textrm{imag}}$ on the imaginary branch
($j_{\textrm{max}}$ is the number of points on the forward branch of the 
Keldysh contour).  Using this scheme, a Dirac delta function is approximated by
\begin{equation}
\delta_c(\bar t_j-\bar t_{j^\prime})=\frac{1}{W_j}\delta_{j,j^\prime}.
\label{eq: delta}
\end{equation}

The evaluation of a discrete approximation to the determinant of the continuous
operator is
now a straightforward procedure~\cite{brandt_urbanek}.  First we note that
\begin{equation}
\textrm{Det}_c (1+M)=\exp (\textrm{Tr}_c[\ln \{1+M\}])=\exp(\sum_n\frac{1}{n}
\textrm{Tr}_c M^n ),
\label{eq: determinant}
\end{equation}
is a relation relating the determinant to the trace of a series
of powers of the matrix
$M$.  The symbol $\textrm{Tr}_c$ denotes the trace of a matrix operator over the
Keldysh contour, and it satisfies
\begin{equation}
\textrm{Tr}_c M = \int_c d\bar t M(\bar t, \bar t)=\sum_i W_i M(t_i,t_i).
\label{eq: trace}
\end{equation}
Hence the trace of the powers of $M$ becomes
\begin{equation}
\textrm{Tr}_c M^n=\sum_{i_1...i_n}W_{i_1}...W_{i_n}M(t_{i_1},t_{i_2})...
M(t_{i_n},t_{i_1}).
\label{eq: trace2}
\end{equation}
Now we define a new discrete matrix to satisfy $\bar M(t_i,t_j)=W_iM(t_i,t_j)$.
Then we find the trace in Eq.~(\ref{eq: trace2}) becomes
\begin{equation}
\textrm{Tr}_c M^n=\sum_{i_1...i_n}\bar M(t_{i_1},t_{i_2})\bar M(t_{i_2},t_{i_3})
...\bar M(t_{i_n},t_{i_1})=\textrm{Tr}\bar M^n,
\label{eq: trace3}
\end{equation}
and leads to the final formula for the determinant
\begin{equation}
\textrm{Det}_c(1+M)=\textrm{Det}(1+\bar M),
\label{eq: determinant2}
\end{equation}
which approximates the determinant of the continuous matrix operator defined 
over the Keldysh contour by a matrix determinant of the discrete matrix
$1+\bar M$.  Hence, for each value of $t$ that we wish to calculate the
$f$-electron Green's function, we must first generate the corresponding
matrix $1+\bar M$ for the Keldysh contour that runs out to time $t$ and
then take its determinant.  Since the matrix can be generated solely from
the parameters $U$, $t$, $T$, and $\lambda(\omega)$, this algorithm is easily
parallelized. We first generate
the function $\lambda(\omega)$ on a discrete real frequency grid [by solving 
for the conduction electron Green's function, Eq.~(\ref{eq: cubic})] 
on the master node, and then send that data to
all of the slave nodes of the parallel process.  Each slave process calculates 
the relevant matrix (which is a general complex matrix, with no special 
symmetries or properties), diagonalizes the matrix to find its eigenvalues,
and then computes the determinant by taking a product of the eigenvalues. This
is then sent back to the master who computes $G_f^>(t)$ from 
Eq.~(\ref{eq: g_f_t}) and sends a new value of $t$ to the slave to continue
the computation.  The algorithm has essentially a linear scale up in
the parallelization, and it is quite efficient, since the limiting step
is the diagonalization of the matrix, which is optimized by the local
implementation of LAPACK and BLAS on the given parallel computer. We perform
the majority of our calculations on a Cray T3E, which limits the matrix sizes
to approximately $2100\times 2100$ on a 256~MEG node, and we can increase
the matrix size slightly when working on higher memory nodes, but the 
diagonalization time then becomes a limiter, if it takes longer than the
queue limits for those nodes on a given machine.

We fix the grid spacing on the imaginary axis, since we find the results
are not too sensitive to changes of the step size there, and the value
$\Delta t_{\textrm{imag}}=0.05$ is sufficient for our purposes.  On the
real time axis, we generally take $\Delta t_{\textrm{real}}$ to vary from
0.1 down to 0.0125.  But since the calculations at different values of $t$
are completely independent of one another, we do not need to use the
same grid spacing of the $t$-values for which we compute $G_f^>(t)$.  
We find that choosing a real time axis spacing of 0.2 or 0.1 
[for generating the data for $G_f^>(t)$] is usually
sufficiently accurate.  We perform an Akuba shape preserving spline and evaluate
the splined Green's function on a grid of size 0.01 or 0.005 before evaluating
the cosine Fourier transform in Eq.~(\ref{eq: f_spect_final}). This allows
the accuracy to improve for larger frequency values, without producing a 
significant increase in computational time.  Finally, the results of the
Fourier transform, especially at small frequencies, depend on the cutoff
or maximal time value where $G_f^>(t)$ is evaluated.  Since the Green's
function decays at large times, imposing a cutoff is like replacing the
Green's function by 0 for times larger than the cutoff.  We find that this
usually
causes no significant errors to the calculations when the maximum value of the
Green's function is less than approximately
$10^{-4}$ to $10^{-5}$ in magnitude at the
point where the cutoff is imposed.

\section{Numerical Results}

When $U=0$, the system is noninteracting, and the $f$-electron DOS is
a delta function for all temperatures.  When $U$ increases, the $f$-electron
DOS broadens into a regular
function and picks up $T$-dependence (surprisingly, the
conduction electron DOS is always independent of 
temperature~\cite{vandongen}). For small values
of $U$, we expect the low-temperature $f$-electron Green's function to be
a sharply peaked function with unit weight. The Fourier transform of this
function 
to the real axis, will then be an exponentially decaying function, with
a slower decay for a more sharply peaked function.  Hence, we expect the
greater Green's function to have an exponential tail for large times.  At
small times, the Green's function approaches $\rho_f-1$ for all temperatures, 
and 
the curves for different temperatures start to separate only for larger times.

\begin{figure}
\epsfxsize=3.1in
\centerline{\epsffile{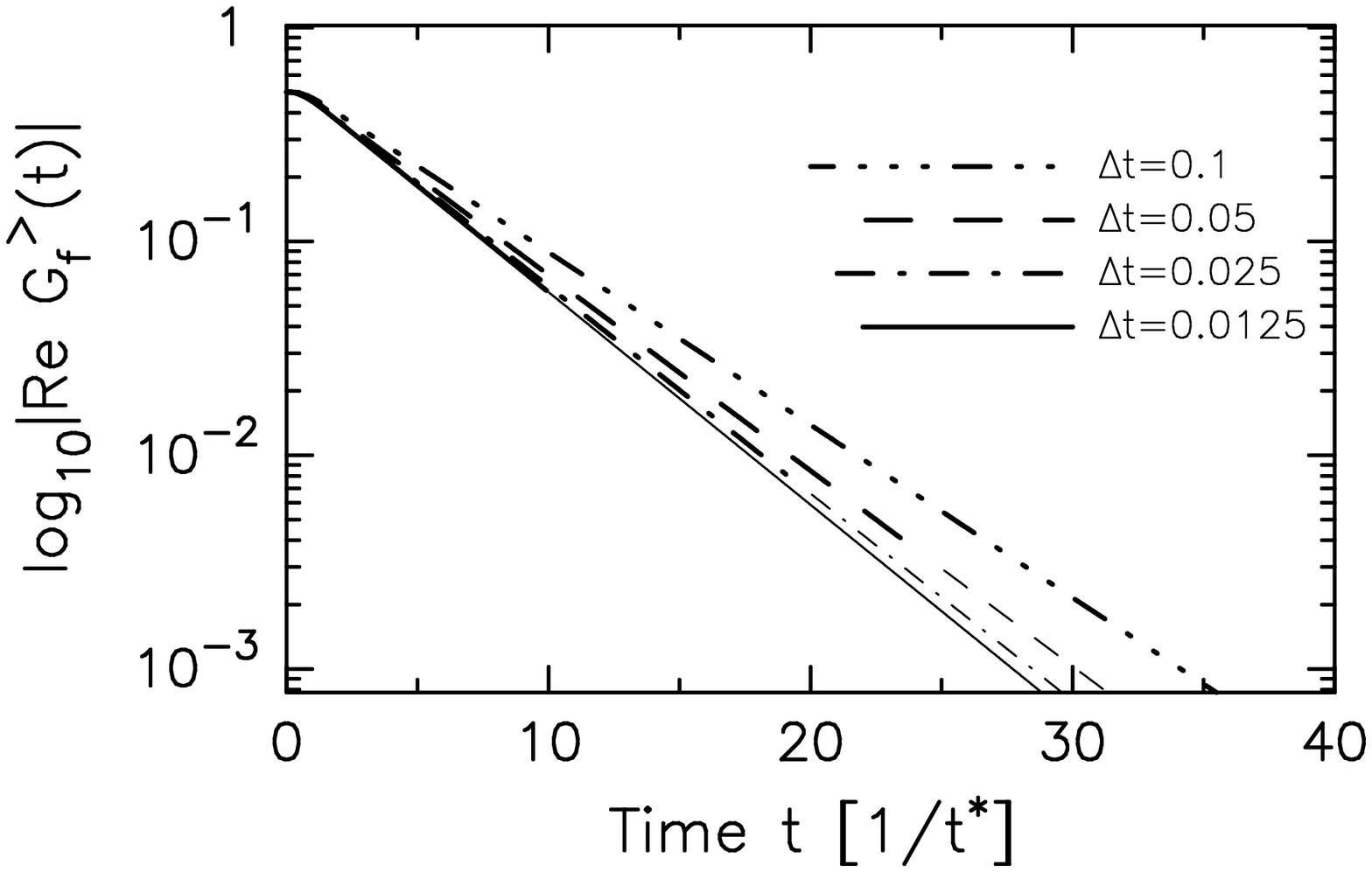}}
\caption{\label{fig: lng-5}Logarithm of the absolute value of $G_f^>(t)$ for
$U=1$ and $T=5$.  We plot results for $\Delta t_{\textrm{real}}=0.1$, 0.05,
0.025, and 0.0125.  Note how the Green's function has a simple exponential decay
at large times, and how the decay constant depends on the
discretization size. The thick curves are the calculated data, and the thin
curves are the extrapolated results.}
\end{figure}

We illustrate the output of our calculations for the case $U=1$ and $T=5$
in Fig.~\ref{fig: lng-5}.  We plot the logarithm of the absolute value
of the real part of $G_f(t)$ for four different choices of 
$\Delta t_{\textrm{real}}$.  We find, in all cases, that the Green's function 
has an exponentially decaying behavior at large times, so we append an 
exponential function out to large times, in order to have the Green's 
function smaller than $10^{-4}$ at the maximal $t$ cutoff. The extrapolated
curves are represented by the thin lines.  Note how there is a clear
dependence of the Green's function on the step size taken along the real axis.
Since this is a semi-log plot, it suggests that one extrapolate the logarithm of
$G_f^>(t)$ to determine the $\Delta t_{\textrm{real}}\rightarrow 0$ limit. But
this procedure becomes problematic once the Green's function crosses zero,
or has oscillatory behavior in the tails, so we do not carry out such
a procedure here (note one might have expected $G_f^>(t)$ to have a 
quadratic dependence on $\Delta t_{\textrm{real}}$ due to a Trotter formula,
but that does not apply here, since there is no simple Trotter breakup of
the continuous matrix operator we are computing the determinant of).

\begin{figure}
\epsfxsize=3.1in
\centerline{\epsffile{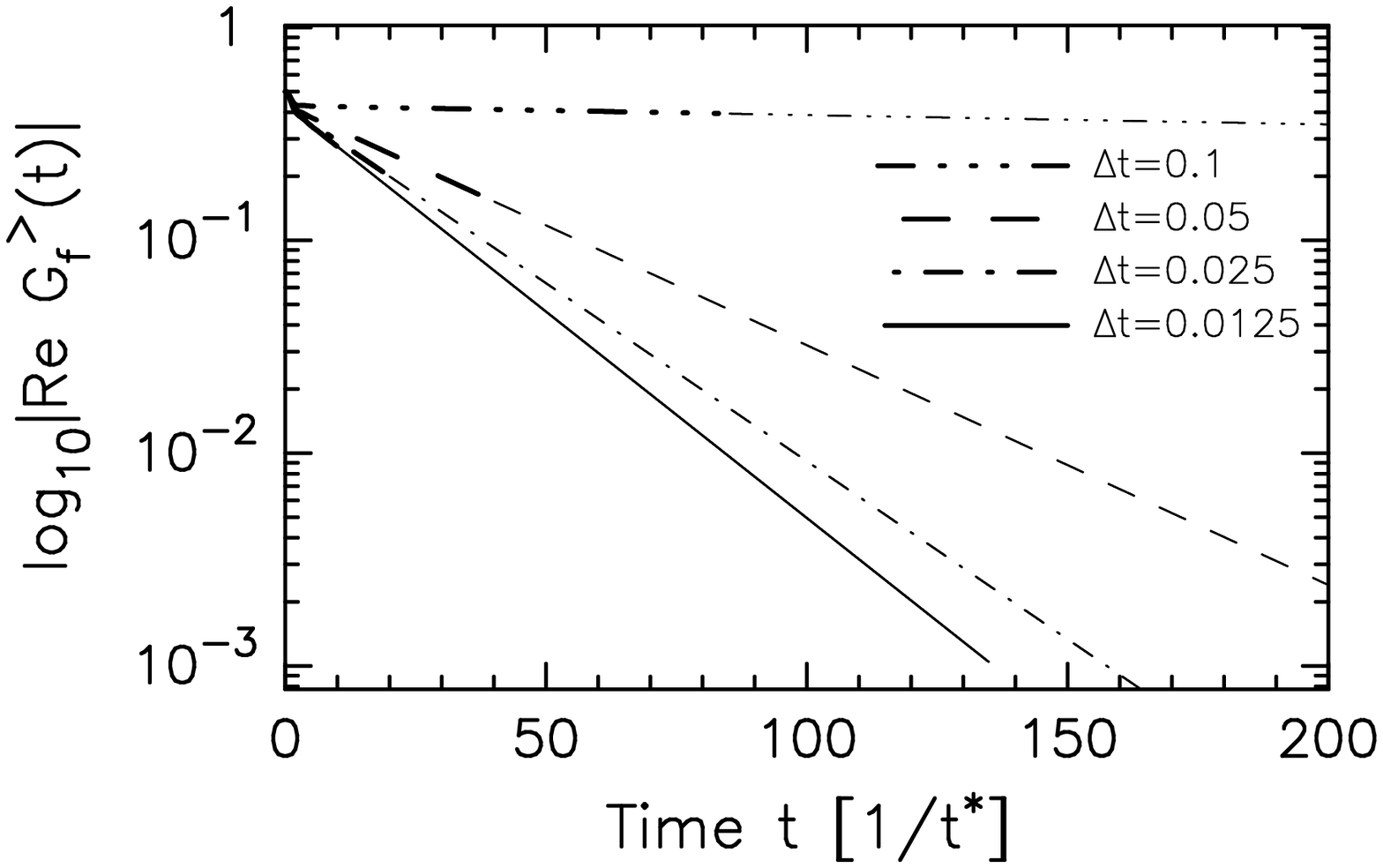}}
\caption{\label{fig: lng-0.15}Logarithm of the absolute value of $G_f^>(t)$ for
$U=1$ and $T=0.15$.  We plot results for $\Delta t_{\textrm{real}}=0.1$, 0.05,
0.025, and 0.0125.  Note how the Green's function has a simple exponential decay
at large times, and how the decay constant depends strongly on the 
discretization size. The thick curves are the calculated data, and the thin
curves are the extrapolated results.}
\end{figure}

Next we examine the same set of parameters, but at lower temperature $T=0.15$
in Fig.~\ref{fig: lng-0.15}.  Here the dependence on the discretization
size is much stronger, with the exponential decay quite slow for the
largest $\Delta t_{\textrm{real}}$.  This shows that the discretization size
needs to be reduced as $T$ is reduced, making lower temperature calculations
much more difficult than higher temperature.  Also, the maximal cutoff in
time needs to be pushed farther out, unless one can append an extrapolated
functional form (as we do here) for large times.

The next step is to perform the Fourier transform (after splining the time
data) as in Eq.~(\ref{eq: f_spect_final}). We can then take the data for the
DOS for different $\Delta t_{\textrm{real}}$ 
values and try to perform a pointwise
(in $\omega$) extrapolation down to $\Delta t_{\textrm{real}}=0$.  Since we
do not know how the curves depend on $\Delta t_{\textrm{real}}$, we use
an $n$-point Lagrange interpolation formula, which is a linear extrapolation
for $n=2$, a quadratic extrapolation for $n=3$ and a cubic extrapolation
for $n=4$.  By checking the different sum rules and the values of the Green's
function at the Matsubara frequencies, we can examine the accuracy of different
extrapolation schemes.  Sometimes it is better to use all of the data and
a large $n$ Lagrange extrapolation scheme.  Other times, the large step-size
error is so big, that that data is not trustworthy, and it is more accurate to 
use an extrapolation with just the smaller discretization sizes (usually a
linear extrapolation method with the smallest two values of
$\Delta t_{\textrm{real}}$ is used).  We call this extrapolation technique
the $\delta$-extrapolation.

Often, we find that the exact value for the Green's function at the lowest
Matsubara frequency lies in between the value for the smallest 
$\Delta t_{\textrm{real}}$ and the value generated from the 
$\delta$-extrapolation.  In this case, we usually can improve the DOS if we 
perform a second extrapolation, averaging those two DOS to produce the correct
value for $G_f(i\omega_0)$.  We call this extrapolation scheme 
Matsubara-extrapolation.

\begin{figure}
\epsfxsize=3.1in
\centerline{\epsffile{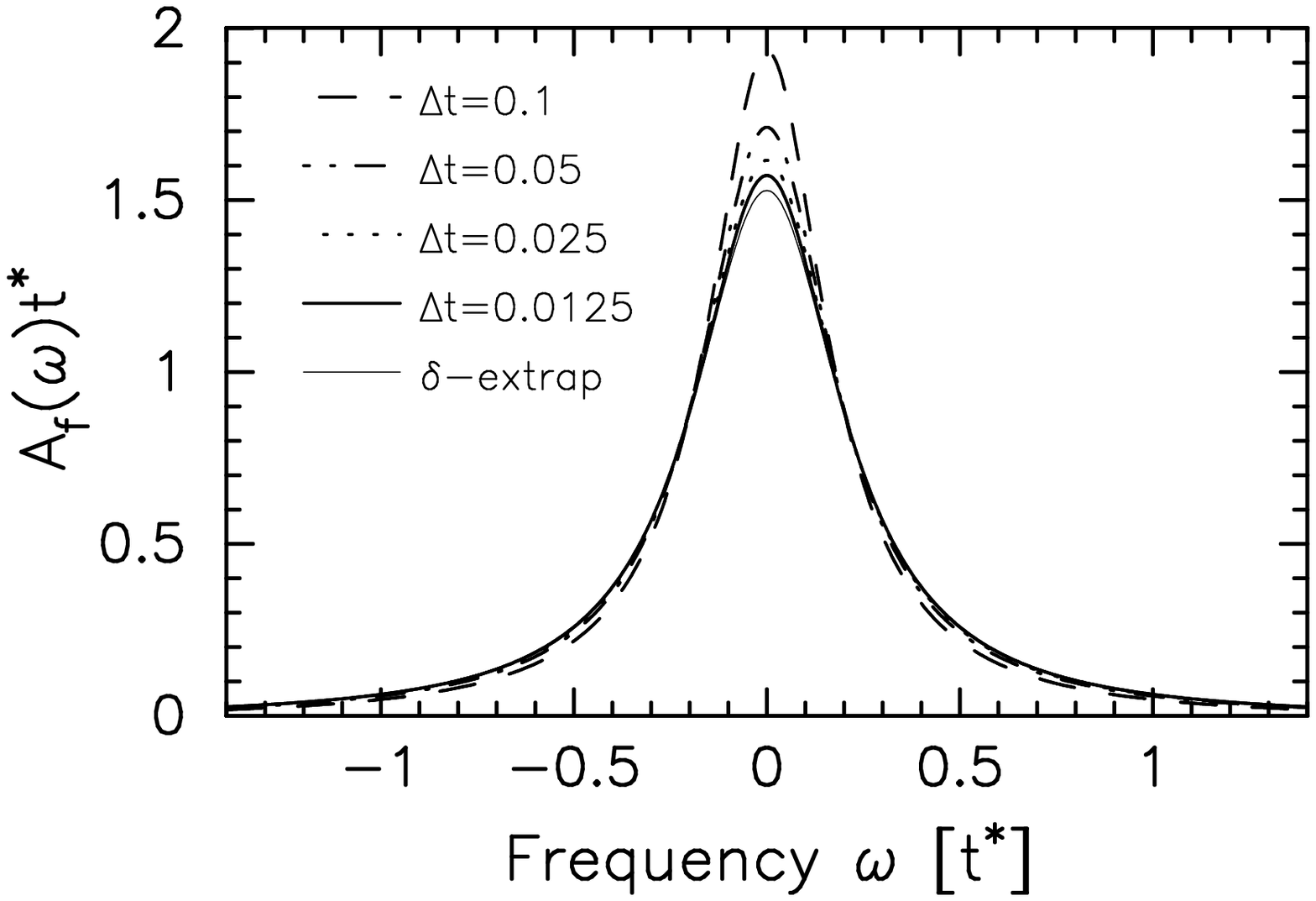}}
\caption{\label{fig: dos_t=5}
DOS for $U=1$ and $T=5$ and different values of $\Delta t_{\textrm{real}}$.
Also plotted is the result of the $\delta$-extrapolation.  One can see clearly
that the DOS is converging to a well-defined limit as the discretization
error is reduced.
}
\end{figure}

To illustrate how the extrapolation procedures work, we first examine the
high-temperature case $T=5$.  The results for the DOS for the four
different discretization sizes and for the $\delta$-extrapolation is
shown in Fig.~\ref{fig: dos_t=5}. It is apparent from the figure that the 
DOS broadens and the peak is reduced as the discretization size is made
smaller. The $\delta$ extrapolation uses the two smallest values of
$\Delta t_{\textrm{real}}$ and a linear extrapolation (we found that gave
the most accurate results).  A summary of the moment sum rules and the spectral
formula for the Matsubara Green's functions is given in 
Table~\ref{table: U=1_T=5}.  It is clear from that data that a systematic 
extrapolation is possible, and that the final result is highly accurate for 
the DOS (errors are less than 0.1\% for the first moment, 0.03\% for the 
second moment, and 0.003\% for the lowest Matsubara frequency). It is hard to 
judge the absolute accuracy of the DOS from these integrated sum rules,
but as a general rule, if the moment sum rules are accurate to better than
1\% and the Matsubara frequency Green's functions are accurate to better
than 0.1\%, then the DOS has an absolute accuracy that is probably
better than a few percent for most frequencies, except those near the tails,
where the DOS is small and may even go negative.

\begin{table}[h]
\caption{Summary of moment sum rules and the Matsubara Green's function
for the case $U=1$ and $T=5$. The first four rows are different values of
$\Delta t_{\textrm{real}}$. The last row is the exact result.
Shift is the value added to the DOS to satisfy
the sum rule in EQ.~(\ref{eq: moment1}).
\label{table: U=1_T=5}}
\begin{tabular}{llll|l}
Case&Eq.~(\ref{eq: moment2})&Eq.~(\ref{eq: moment3})&$G_f(i\omega_0)$&shift\\
\hline
$0.1$&$-0.00787$&0.15805&$-0.063667$&0.00484\\
$0.05$&$-0.00991$&0.19933&$-0.063660$&0.00500\\
$0.025$&$-0.01087$&0.21863&$-0.063628$&0.00253\\
$0.0125$&$-0.01137$&0.22873&$-0.063623$&0.00125\\
$\delta$-extrap.&$-0.01241$&0.25007&$-0.063600$&$-10^{-6}$\\
\hline
Exact&$-0.01240$&0.25&$-0.063598$&0.0\\
\hline
\end{tabular}
\end{table}

\begin{figure}
\epsfxsize=3.1in
\centerline{\epsffile{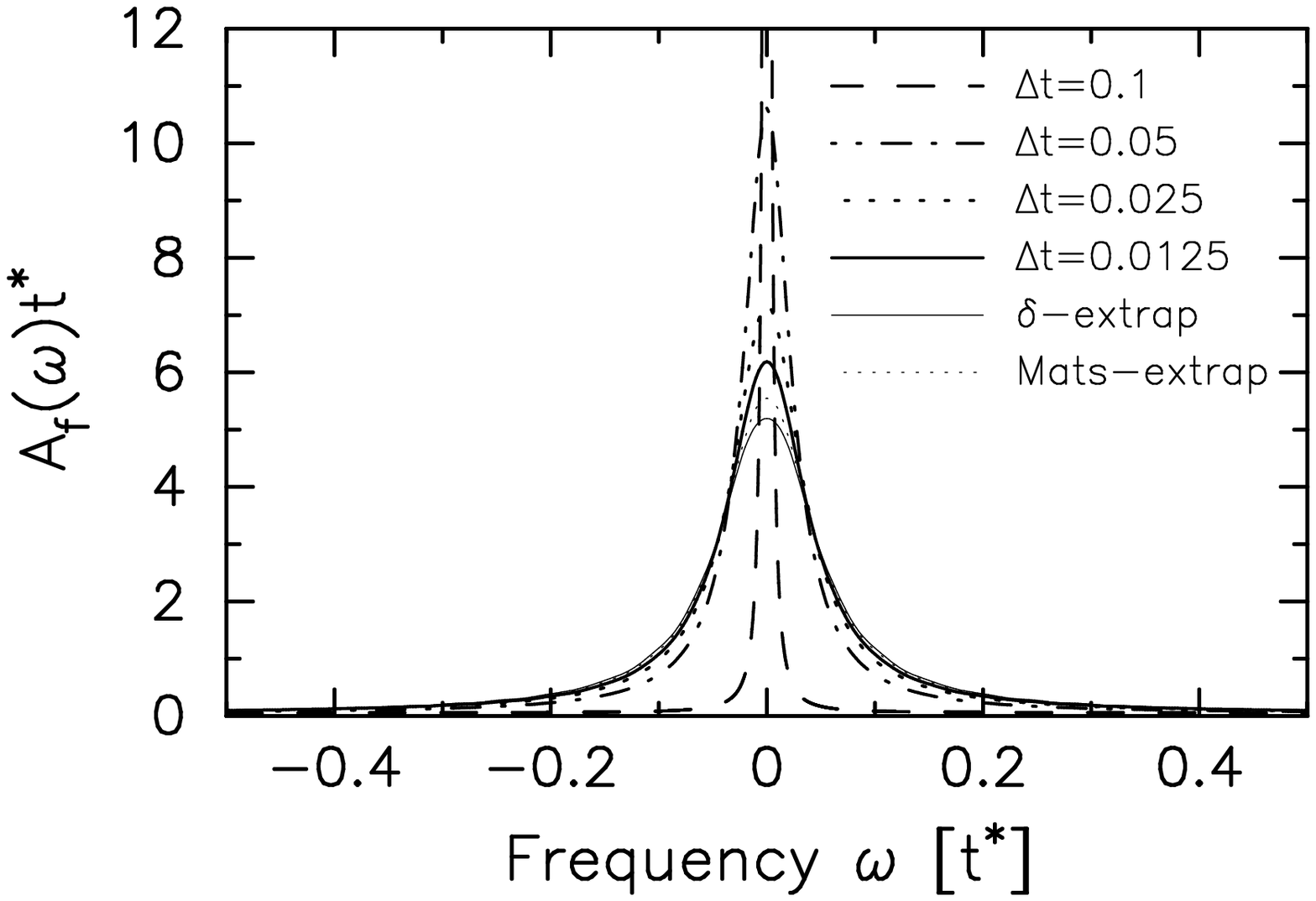}}
\caption{\label{fig: dos_t=0.15}
DOS for $U=1$ and $T=0.15$ and different values of $\Delta t_{\textrm{real}}$.
Also plotted is the result of the $\delta$-extrapolation and the 
Matsubara-extrapolation.  One can see
that the DOS is converging to a well-defined limit as the discretization
error is reduced, but the accuracy is significantly reduced relative to
the higher-temperature case in Fig.~\ref{fig: dos_t=5}.
}
\end{figure}

We next examine the low-temperature case of $U=1$ and $T=0.15$ in 
Fig.~\ref{fig: dos_t=0.15}.  Note how the large $\Delta t_{\textrm{real}}$
case has an extremely narrow DOS, and how the DOS broadens dramatically
as the step size is reduced.  We are able to make this reduction, since
we can append the exponential tail out to large times, which allows
us to perform the relevant Fourier transforms.  We plot two extrapolation
techniques: the $\delta$-extrapolation, using a linear extrapolation with the
two smallest step sizes (since that is the most accurate), and the 
Matsubara-extrapolation procedure, which averages the 
$\Delta t_{\textrm{real}}=0.0125$ DOS with the $\delta$-extrapolated DOS in
order to properly reproduce the lowest Matsubara frequency Green's function.
Since the $\delta$-extrapolated result overshoots the correct answer, this 
averaging procedure significantly enhances the results.  Even so, one can 
see from Table~\ref{table: U=1_T=0.15} that the errors are much larger
at low temperature than at high temperature (this was already apparent from
Fig.~\ref{fig: lng-0.15}).  The error in the moment from Eq.~(\ref{eq: moment2})
is 1\%, the moment from Eq.~(\ref{eq: moment3}) is 1.5\% and 
for the Matsubara frequency Green's function $G_f(i\omega_1)$ is
$0.07\%$.

\begin{table}[h]
\caption{Summary of moment sum rules and the Matsubara Green's function
for the case $U=1$ and $T=0.15$. The first four rows are different values of
$\Delta t_{\textrm{real}}$. The last row is the exact result.
\label{table: U=1_T=0.15}}
\begin{tabular}{lllll}
Case&Eq.~(\ref{eq: moment2})&Eq.~(\ref{eq: moment3})&$G_f(i\omega_0)$&
$G_f(i\omega_1)$\\
\hline
0.1&$-0.05361$&0.12596&$-1.94404$&$-0.6832$\\
0.05&$-0.08145$&0.20057&$-1.855512$&$-0.6742$\\
0.025&$-0.09030$&0.21792&$-1.81501$&$-0.6694$\\
0.0125&$-0.09760$&0.24410&$-1.79435$&$-0.6668$\\
$\delta$-extrap.&$-0.10307$&0.25936&$-1.77368$&$-0.6642$\\
M-extrap.&$-0.10112$&0.25390&$-1.78105$&$-0.6651$\\
\hline
Exact&$-0.10217$&0.25&$-1.78106$&$-0.6656$\\
\hline
\end{tabular}
\end{table}

\begin{figure}
\epsfxsize=3.1in
\centerline{\epsffile{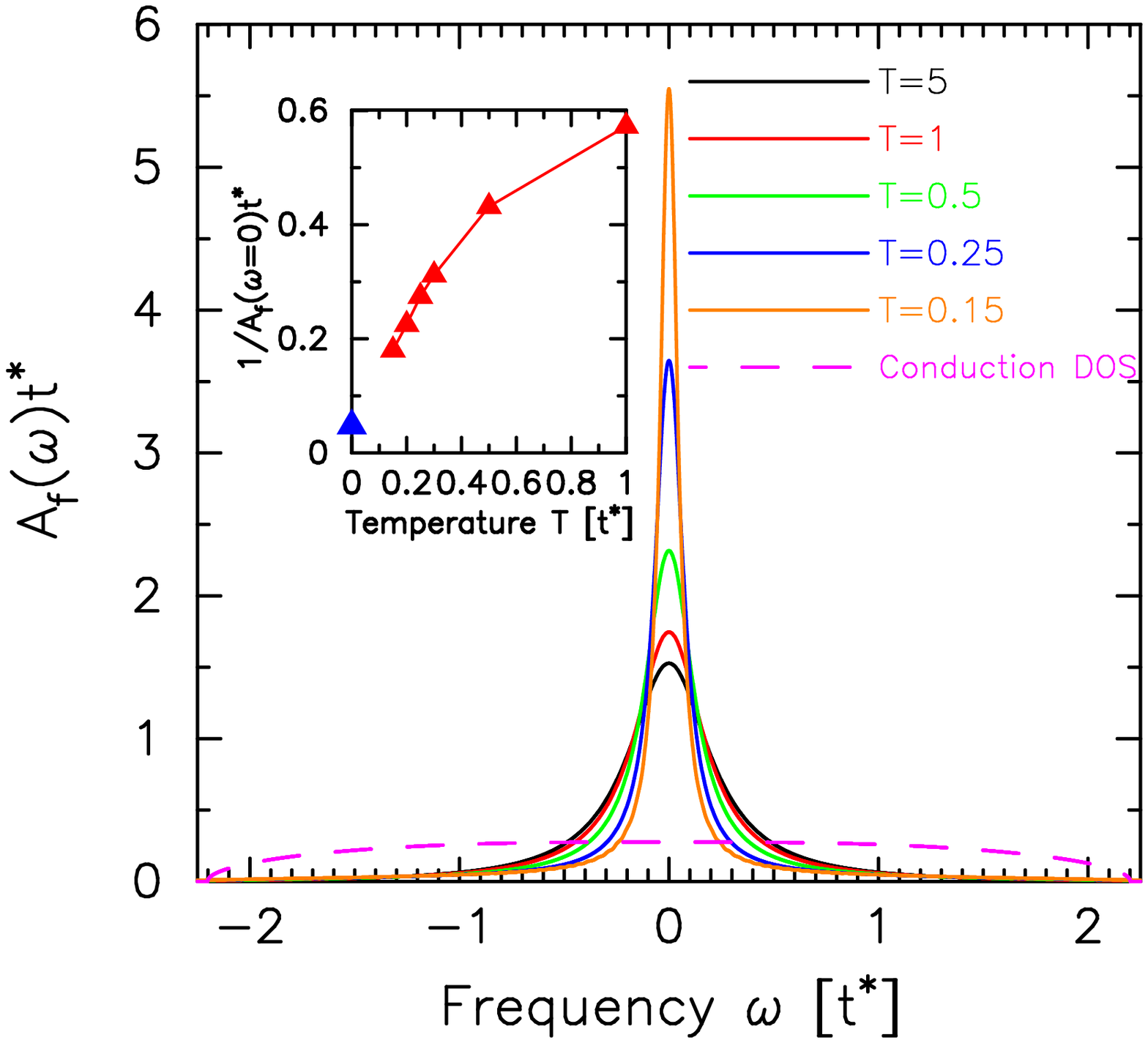}}
\caption{\label{fig: u=1_summary}
Summary plot for the $f$-electron DOS for $U=1$ and a number of different
temperatures.  The magenta dashed line is the conduction electron DOS (which is
independent of temperature).  Inset is a plot of $1/A_f(\omega=0)$ versus $T$,
to extrapolate the DOS down to $T=0$.  We predict that the DOS grows to a peak
height of about 21 at $T=0$ (blue triangle in inset).
}
\end{figure}

Hence these calculations become significantly more challenging as the
temperature is lowered.  To understand the thermal evolution of the
$f$-electron DOS, we summarize data collected for a number of different
temperatures in Fig.~\ref{fig: u=1_summary}.  In addition, we plot
the temperature-independent conduction electron DOS with the magenta 
dashed line.
Note how the conduction electron DOS is rather broad, but the $f$-electron
DOS becomes sharply peaked at low temperature.  By scaling the results
to $T=0$, we conjecture that the maximum of the $f$-electron DOS 
approaches 21 as
$T\rightarrow 0$.  As the DOS becomes more sharply peaked, we need to
have a transfer of some spectral weight to larger energy as well ($|\omega|>
1.5$), since the
second moment sum rule [in Eq.~(\ref{eq: moment3})] 
is independent of temperature, and as the peak grows in
height, it contributes less to that sum rule.  There is an interesting contrast
in the DOS of the two different particles.  The conduction electron DOS is broad
and does not evolve with temperature, while the $f$-electron DOS has strong
temperature dependence becoming sharply peaked at low temperature and
weak coupling.  By carefully performing extrapolations of the exponential
decay of the greater Green's function at large time and of the discretization
error of the DOS, we can produce accurate results for the $f$-electron
DOS.  The calculational needs
can easily go beyond computational resources at low temperature, though.

Next we focus on the strong-coupling regime with $U=5$.  In this case, the
conduction electron DOS has a correlation induced gap from the Mott transition,
which occurs at $U=2$.  When the DOS develops a gap at low $\omega$, the 
greater Green's function has a strong oscillatory component at large times.
Unfortunately there does not appear to be any simple rule that could
be used to append an extrapolated result to the Green's function at large
time.  This greatly reduces the ability to calculate accurate results
at low temperature, because the discretization size needs to be small, but
the cutoff in time needs to be large, and this often goes beyond available
computer resources.

\begin{figure}
\epsfxsize=3.1in
\centerline{\epsffile{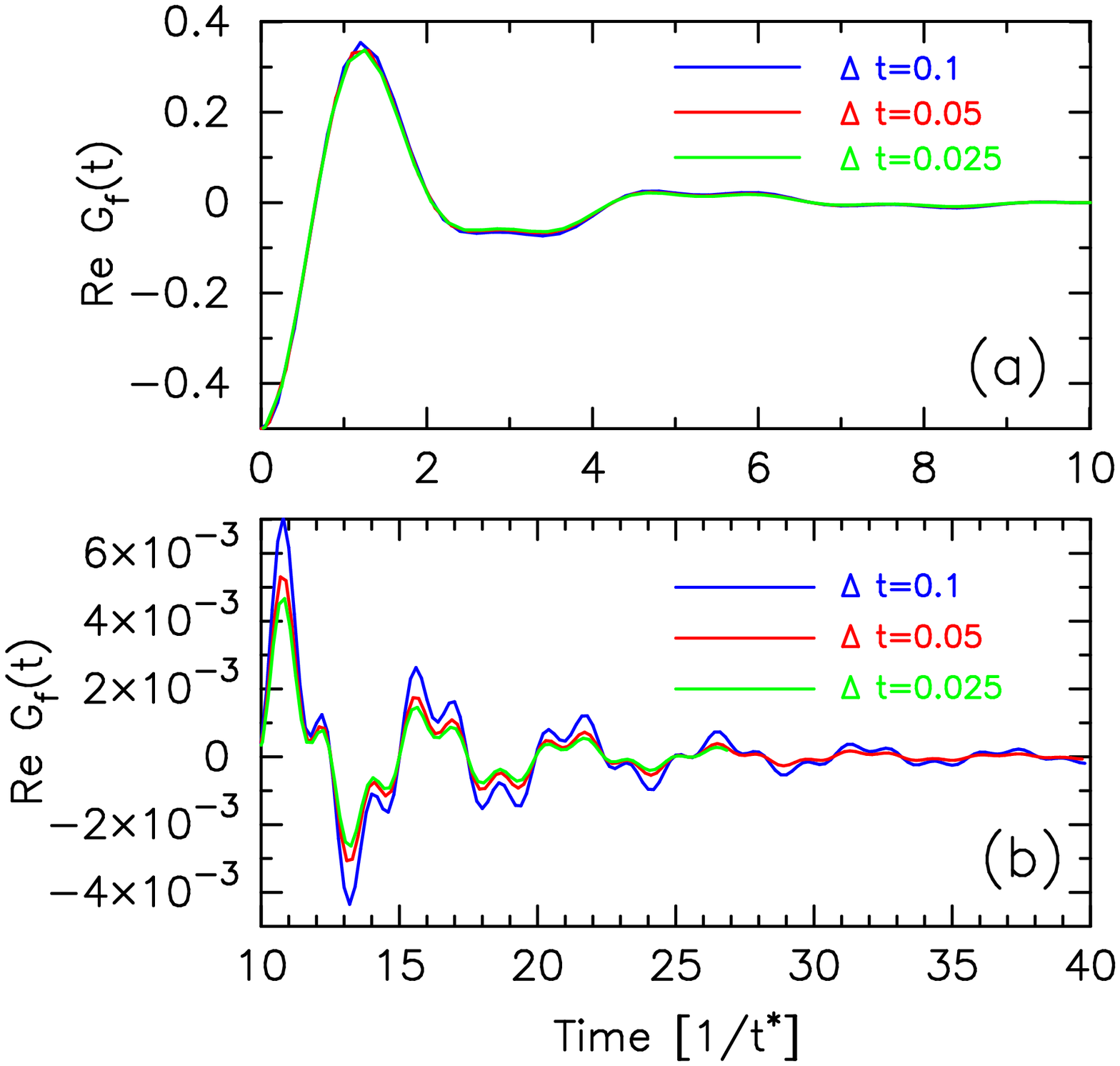}}
\caption{\label{fig: Gf_t_u=5}
Greater Green's function for $U=5$ and $T=1$ for different 
$\Delta t_{\textrm{real}}$.  Panel (a) is the short-time results and panel
(b) is the longer-time results.  Note how the curves are virtually identical for
short times, but there is a systematic reduction in the amplitude of the 
oscillations as $\Delta t_{\textrm{real}}$ is reduced.
}
\end{figure}

In Fig.~\ref{fig: Gf_t_u=5}, we plot the greater Green's function
versus time for $T=1$.  In panel (a), the short-time results are shown---note
how the curves for different discretization size lie almost on top of
each other. In panel (b), we show the larger time results (the smallest
discretization size [green curve]
only goes to $t=27$).  What is interesting to note
is that the amplitude of the oscillations is reduced as the discretization
size is reduced, but the period remains essentially the same.  This might
suggest to try to determine the amplitude reduction factor for the limit
$\Delta t_{\textrm{real}}\rightarrow 0$, and apply that to the curves
for finite $\Delta t_{\textrm{real}}$ to extrapolate to the exact 
result.  The problem is that such a scheme does not work well, as it
tends to generate an unphysical peak in the low-frequency DOS, and it produces
worse agreement for both the moments and the Matsubara Green's functions,
so we don't discuss that scheme further.

\begin{figure}
\epsfxsize=3.1in
\centerline{\epsffile{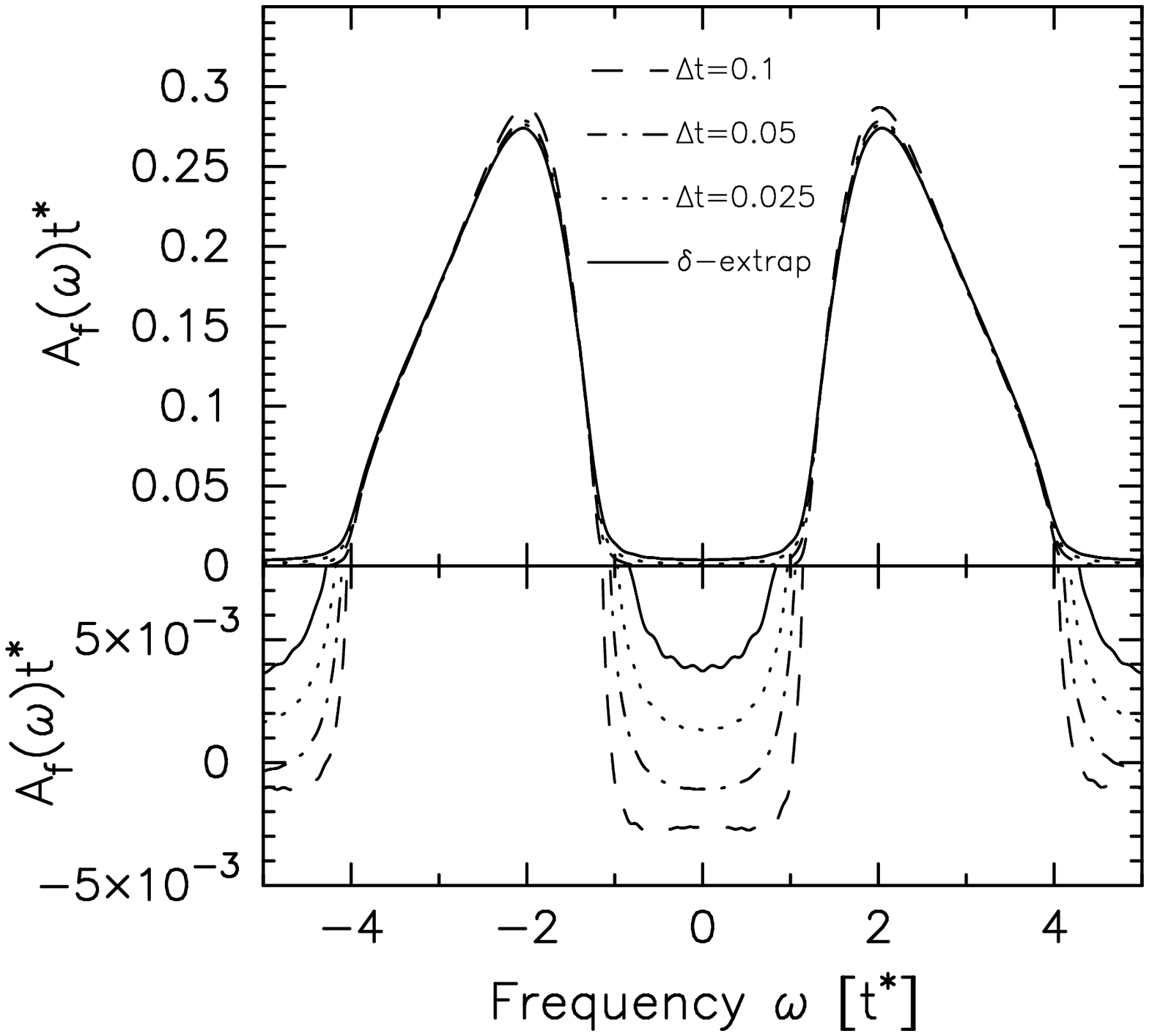}}
\caption{\label{fig: u=5_t=1}
DOS for $U=5$ and $T=1$.  Three different values of $\Delta t_{\textrm{real}}$
and the $\delta$ extrapolation are shown. In the lower panel, the vertical axis
of the figure is blown up to highlight
that the DOS in the gap region is negative for large discretizations
and becomes positive only as the discretization is reduced.
}
\end{figure}

We are thus left with using the same $\delta$ and Matsubara extrapolation 
schemes that we used for the small-$U$ case.  At high temperature, everything
works out reasonably well, as can be seen in Table~\ref{table: u=5_t=5}. But
we do not achieve anywhere near as high an accuracy as for small $U$.  When we
go to lower temperatures, the accuracy becomes worse, and there is limited
improvement from the extrapolation schemes.  This occurs because when the
discretization size is too large, the DOS becomes negative in the gap region,
while when the size is small enough to correct the DOS in the gap region,
it suffers from Gibb's oscillations due to the sharp cutoff in time,
since the cutoff is not large enough for an accurate Fourier transform.
These results are summarized in Fig.~\ref{fig: u=5_t=1} and 
Table~\ref{table: u=5_t=1}. Note that the sum rules do not change much as
the DOS varies in the gap region, because the overall DOS is small and 
because we multiply by powers of $\omega$ which suppress the weight in the
integral from the gap region.

\begin{table}[h]
\caption{Summary of moment sum rules and the Matsubara Green's function
for the case $U=5$ and $T=5$. The first two rows are different values of
$\Delta t_{\textrm{real}}$. The last row is the exact result.
Shift is the value added to the DOS to satisfy
the sum rule in EQ.~(\ref{eq: moment1}).
\label{table: u=5_t=5}}
\begin{tabular}{llll|l}
Case&Eq.~(\ref{eq: moment2})&Eq.~(\ref{eq: moment3})&$G_f(i\omega_0)$&shift\\
\hline
$0.1$&$-0.30191$&6.19393&$-0.062162$&0.00486\\
$0.05$&$-0.30299$&6.21921&$-0.062147$&0.00502\\
$\delta$-extrap.&$-0.30396$&6.24237&$-0.062120$&$0.00010$\\
\hline
Exact&$-0.30422$&6.25&$-0.062101$&0.0\\
\hline
\end{tabular}
\end{table}

\begin{table}[h]
\caption{Summary of moment sum rules and the Matsubara Green's function
for the case $U=5$ and $T=1$. The first three rows are different values of
$\Delta t_{\textrm{real}}$. The last row is the exact result.
Shift is the value added to the DOS to satisfy
the sum rule in EQ.~(\ref{eq: moment1}).
\label{table: u=5_t=1}}
\begin{tabular}{llll|l}
Case&Eq.~(\ref{eq: moment2})&Eq.~(\ref{eq: moment3})&$G_f(i\omega_0)$&shift\\
\hline
$0.1$&$-1.01562$&6.28460&$-0.203556$&0.00504\\
$0.05$&$-1.01334$&6.30267&$-0.203146$&0.00507\\
$0.025$&$-1.01012$&6.29576&$-0.202731$&0.00504\\
$\delta$-extrap.&$-1.00954$&6.30842&$-0.202735$&$-0.00315$\\
\hline
Exact&$-1.00177$&6.25&$-0.203916$&0.0\\
\hline
\end{tabular}
\end{table}

\begin{figure}
\epsfxsize=3.1in
\centerline{\epsffile{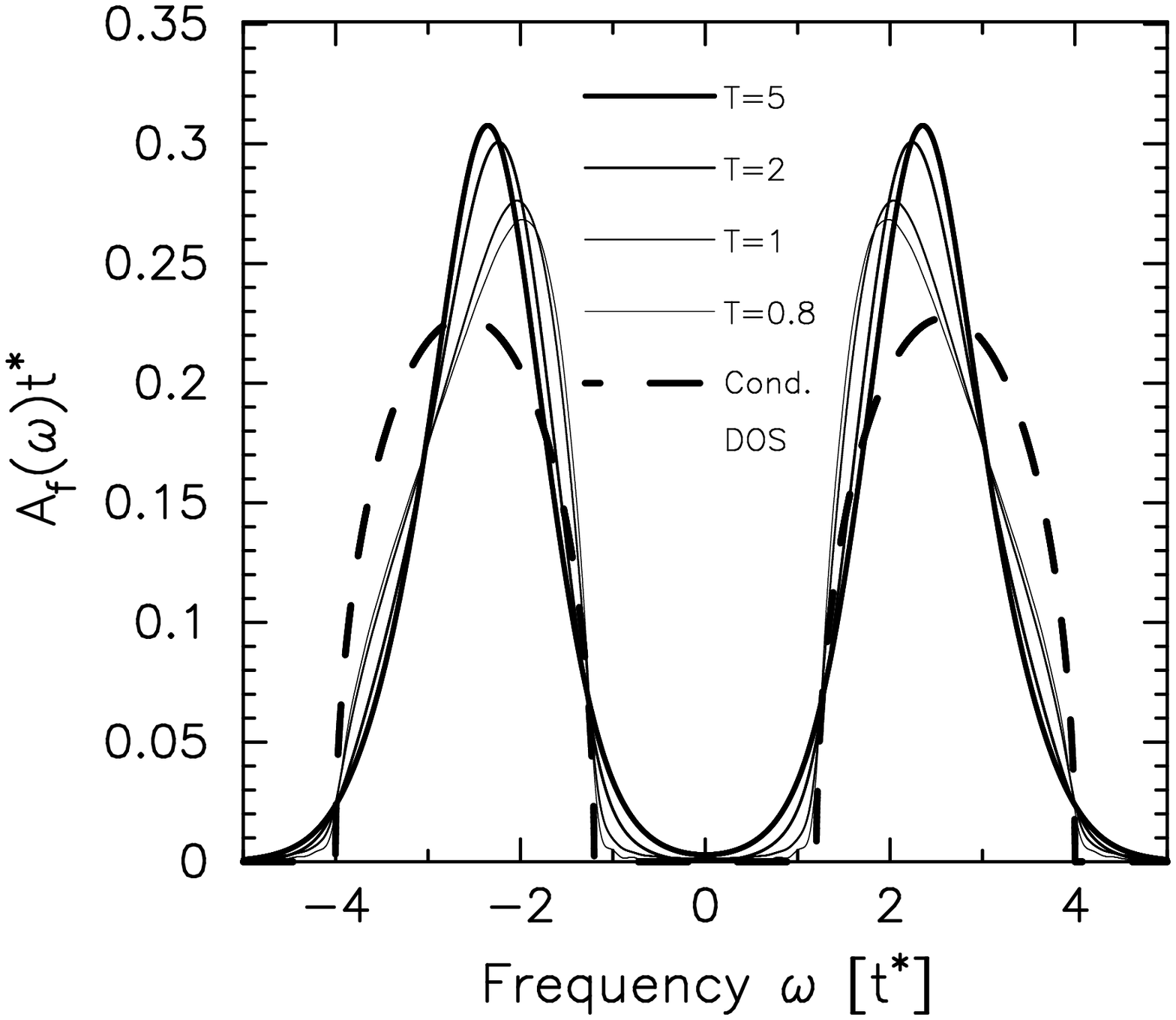}}
\caption{\label{fig: u=5_summary}
Summary DOS for $U=5$. The data for $T=5$ comes from the $\delta$-extrapolation,
for $T=2$ from $\Delta t_{\textrm{real}}=0.05$, and the lower temperatures
have $\Delta t_{\textrm{real}}=0.025$.  We also include the conduction
electron DOS with the dashed line.  Note how the subgap DOS for the
$f$-electron decreases as $T$ is lowered, and how the DOS moves more towards
the band edge as $T$ is reduced.  We also see the large $|\omega|$ DOS start
to pinch off at the conduction electron DOS band edge
as $T$ is lowered. We expect
the bandwidths of both the conduction and $f$ electrons to be equal at
$T=0$. The oscillations apparent in the $T=0.8$ data are an artifact of
a time-domain cutoff that is too short.
}
\end{figure}

Finally, a summary of the DOS data for $U=5$ is plotted in 
Fig.~\ref{fig: u=5_summary}.  We see that there is significant subgap
DOS at high temperature, which is reduced as $T$ is lowered.  We also see the
peak in the $f$-electron DOS move towards the band edge as $T$ is lowered.
Finally, it appears that the $f$-electron DOS and the conduction electron DOS 
will both share the same bandwidth at $T=0$.  We are severely limited by how
low we can go in temperature and still maintain positive DOS in the gap region.

\section{Conclusions and Outlook}

In this contribution, we have presented a summary of numerical calculations
employing a nonequilibrium formalism over the so-called Keldysh contour.
The calculations involve determining the determinant of a continuous
matrix operator defined along the Keldysh contour, which is found by
calculating the determinant of a discretized version of the operator, which
is a general
complex matrix, defined differently for each value of time. This formalism
is easily parallelized, is efficient on each node,
and has nearly linear scale-up. We examined the
simplest problem with this numerical algorithm---the $f$-electron spectrum
of the Falicov-Kimball model.  This problem is useful because a number of
sum rules exist, that allow one to determine the accuracy of the 
calculations.

Our results show that one needs to carefully perform an analysis of the
dependence of the solutions on the discretization along the Keldysh 
contour.  Sometimes results can be extrapolated to the continuum limit,
other times, this is not possible.  We also find that these results often
require a finer discretization at lower temperature. We anticipate similar 
numerical issues will arise in other nonequilibrium problems that employ
the same formalism.

These results provide useful benchmarks for more interesting nonequilibrium
problems such as the interaction of a strongly correlated material with
a strong external electromagnetic field or the nonlinear 
response of an inhomogeneous multilayered
device to an external voltage (including a noise analysis of the current).
These latter problems are likely to have more of an applied interest to
the military.  The calculational formalism needs to be generalized for 
each of these cases.  One needs to find a way to self-consistently map an
impurity or cluster problem onto the lattice problem (in the presence of
the time-dependent field), and then perform similar computations along the
Keldysh contour.  The impurity-like problems are similar, but we need to
perform a summation over the lattice wavevector (which can be represented by
a double integral over a generalized joint density of states for each matrix
element of the contour-ordered Green's function) to complete the 
self-consistent algorithm.
Far fewer sum rules will exist to benchmark the results,
so an analysis in terms of scaling with respect to the discretization size
will need to be performed.  One can check the nonequilibrium results in
the linear-response regime with the results of Kubo-formula-based 
approaches, which will provide a stringent test of the quality of the numerics.

\newpage

\section{Acknowledgments}
We acknowledge support of the Office of Naval Research under grant
number N00014-99-1-0328 and from the National Science Foundation under
grant number DMR-0210717.  Supercomputer resources  (Cray T3E) were provided by
the Arctic Region Supercomputer Center (ARSC) and the Engineering Research
and Development Center (ERDC).

\end{document}
